\documentclass[%
 reprint,
 amsmath,amssymb,
 aps,
]{revtex4-2}

\usepackage{graphicx}
\usepackage{dcolumn}
\usepackage{bm}
\usepackage[colorlinks=true, linkcolor=black, citecolor=blue, urlcolor=blue]{hyperref}

\begin{document}

\preprint{APS/123-QED}

\title{Efficient Excited-State Calculations for Molecules Based on Contextual Subspace Method and Symmetry Optimizations}

\author{Qianjun Yao}
\author{He Li}%
\email{helix@seu.edu.cn}
\affiliation{School of Electronic Science and Engineering, Southeast University, Nanjing, Jiangsu}
\date{\today}

\begin{abstract}
Quantum computing methods for excited-state calculations remain underexplored in Noisy Intermediate-Scale Quantum (NISQ) hardware, despite their critical role in photochemistry and material science. Herein, we propose a resource-efficient framework that integrates the contextual subspace (CS) method with the Variational Quantum Deflation (VQD) algorithm to enable systematic excited-state calculations for molecules while reducing qubit requirements. On the basis of the numerical results, we find that it is unproblematic to utilize this combination in calculating the excited state to reduce qubits. Furthermore, we demonstrate that the implementation of a spin-conserving hardware-efficient ansatz, namely the $\mathcal{N}(\theta_x,\theta_y,\theta_z)$ block ansatz, allows exploitation of spin symmetry within the projected subspace, thereby achieving further reductions in computational resource demands. Compared to the commonly used $R_{y}R_{z}$ ansatz, using the $\mathcal{N}(\theta_x,\theta_y,\theta_z)$ ansatz can reduce the number of optimization iterations by up to 3 times at a similar circuit depth.
\end{abstract}

\maketitle


\section{Introduction}

Quantum chemistry is a promising area where NISQ~\cite{preskill2018quantum} computers can first show their potential power and may also benefit a lot from the more interesting fault-tolerant quantum computing era due to resource-required algorithms such as quantum phase estimation~\cite{kitaev1995quantum}. The emergence of the Willow chip~\cite{googlequantumaiandcollaborators2024quantum} is undoubtedly a significant milestone in moving toward the next quantum stage. By being able to handle more complex electronic structures and larger systems, it will surely provide a greater impetus to the fields of quantum chemistry \cite{caesura2025faster}, materials \cite{huang2022simulating}, drug development~\cite{ahmad2016photostability}, and biology~\cite{fedorov2021practical}.

Although being largely underexplored in the context of resource-constrained quantum hardware, excited state plays a vital role in the realm of quantum chemistry, specifically in chemical reactions and photochemical processes. 
By learning the energy distribution and transition mechanisms between the energy eigenstates, one can form the basis for chemical kinetics and spectroscopy. The properties of the excited state can lead to applications related to luminescence, novel functional materials, and so on. Therefore, in-depth studies of excited state can provide a theoretical foundation for quantum chemistry and material science~\cite{beard2019comparative}. 

Recent years have witnessed active debate in the quantum computing community regarding how to leverage existing quantum devices to achieve practical utility. Reducing the resources for the problems that we are interested in, e.g., tapering qubits~\cite{bravyi2017tapering} using $Z_{2}$ symmetry and point group symmetry~\cite{setia2020reducing}, is one way to help from the original point. Although techniques like qubit tapering and contextual subspace methods~\cite{kirby2019contextuality, kirby2021contextual} have successfully reduced qubit counts for ground-state calculations, their extension to excited states remains nascent. Herein, we address this gap by integrating the contextual subspace framework with the VQD~\cite{higgott2019variational} method (namely CS-VQD), enabling systematic excited-state calculations with reduced resource demands. 

Variational Quantum Eigensolver (VQE)~\cite{fedorov2022vqe, peruzzo2014variational, kandala2017hardwareefficient} is an algorithm dedicated to the NISQ era which faces a critical challenge in designing the ansatz, a parameterized quantum circuit. The real quantum advantage of such an algorithm is the implementation of the state in a high dimension vector space, which is classically hard without the resource of superposition and entanglement properties.
The tradeoff between expressibility and accessibility of the ground state often makes it a struggle to choose a hardware-efficient or a problem-inspired one. After showing the performance of CS-VQD using a problem-inspired ansatz, we also demonstrate the feasibility of exploiting the spin symmetry after the Hamiltonian is projected into the contextual subspace using a spin-preserved ansatz. Numerical simulations demonstrate that the use of the spin-preserved ansatz achieves a significant reduction in required optimization iterations compared to the conventional $R_{y}R_{z}$ ansatz~\cite{dcunha2023challenges} .

\section{Contextual subspace for excited state calculation}

We first give a brief introduction to the Contextual Subspace Variational Quantum Eigensolver (CS-VQE)~\cite{kirby2021contextual, weaving2023contextual, ralli2023unitary} based on some original papers. 
For a more detailed and rigorous explanation, see the systematic establishment by Weaving et al.~\cite{weaving2023stabilizer}. 
After giving the foundation knowledge, we show the performance of combining the contextual subspace method with the variational quantum deflation method to calculate the excited state.

From a stabilizer perspective, the contextual subspace projection has some kind of similarity to the qubit-tapering technique~\cite{bravyi2017tapering}. It is shown that qubit tapering can be integrated by a contextual subspace projection. Thus we also present our results with the qubit tapering routine in Sec.~\ref{Sec: numericalresults}.
The concept of quantum contextuality originates from the Bell-Kochen-Specker~\cite{bell1964einstein, bell1966problem} theorem. Kirby and Love first built the condition test for an arbitrary Hamiltonian to determine noncontextuality.
Contextuality subspace VQE aims at solving the ground state of a qubit Hamiltonian. This qubit Hamiltonian is composed of Pauli strings which can be defined as,
\begin{align}
P_{N}:=\otimes_{i=0}^{N-1}\sigma_{p}^{(i)}, \, \sigma_{p}\in\{\sigma_{I},\, \sigma_{x},\, \sigma_{y},\, \sigma_{z}\}, 
\end{align}
where $N$ represents the total qubit number and $\sigma_{p}^{(i)}$ means $\sigma_{p}$ acting on $i^{th}$ qubit.

CS-VQE can reduce the number of qubits in a VQE routine but is different from the tapering qubits technique, which exploits the real physical symmetry in the full Hamiltonian.

The contextual subspace method has been used to find a molecule's electronic ground-state energy through a quantum-classical hybrid algorithm, VQE, and thus has been called CS-VQE. This CS-VQE method divides the electronic Hamiltonian into two disjoint parts, which are named the contextual part and the noncontextual parts. The latter one can be effectively solved using a classical computer which indicates a rough result for the original problem. To gain an accurate enough result we would need to solve the contextual part as a quantum correction. In VQE, the Hamiltonian can be represented as a sum of weighted Pauli strings after fermion-qubit transformation and as described above can then be turned into two components,
\begin{align}
  \label{hamiltonian_qubit}
  H_{\text{qubit}}=H_{\text{c}}+H_{\text{nc}}=\sum_{p_{1}} h_{p_{\text{c}}}P_{\text{c}}+\sum_{p_{2}} h_{p_{\text{nc}}}P_{\text{nc}},
\end{align}
 where $p_{1}, \, p_{2}$  are the number of Pauli terms in the contexual part and noncontextual part, respectively. $h_{p_{\text{c}}}$ and $h_{p_{\text{nc}}}$ are real coefficient to guarantee $H_{\text{qubit}}$ is hermitian. The set of Pauli terms in $H_{\text{nc}}$ is required to be closed under inference within all the Pauli terms in $H_{\text{qubit}}$.

 By solving the non-contextual part $H_{\text{nc}}$, we can first obtain a noncontextual ground state energy $E_{\text{nc}}^{\text{g}}$ by minimizing a classical objective function $\eta(\mathbf{\nu}, \, \mathbf{r})$. The former variable is an assignment vector over some selected qubits and can only be valued $\pm 1$ over these sites. The latter variable $\mathbf{r}$ is also a vector but is restricted to unit length. This unit vector relates to an constructed observable $R(\mathbf{r}):=\sum_{i=1}^{M} r_i C_i$, where $C_{i}$ is a representative operator in all the equivalence classes of commutation divided by the symmetry $\mathcal{S}_{\text{nc}}$ of $H_{\text{nc}}$.
 A symmetry of a $H_{\text{nc}}$ means that 
\begin{align}
S:=\{ P \in \{ P_{\text{nc}} \, | \, [P, \, P_{\text{nc}}]=0\}\}.
\end{align}
Additionally, we can extend this set to a symmetry group $\bar{\mathcal{S}}$. Minimizing $\eta(\mathbf{\nu}, \, \mathbf{r})$ to obtain a ground state turns out to be a constrained optimization problem that is classically tractable.

The $E_{\text{nc}}^{\text{g}}$ is shown to be close to the Hartree-Fock energy, demonstrated by Weaving et al. \cite{weaving2023stabilizer}. Choosing different sets of noncontextual terms in $\{P_{N}\}$ leads to different results of $E_{\text{nc}}^{\text{g}}$. However, it is not always true to choose the best set of $\{P_{\text{nc}}\}$ by minimizing $E_{\text{nc}}^{\text{g}}$ \cite{weaving2023stabilizer}.

Suppose that we have defined the noncontextual set $\{P_{\text{nc}}\}$ and calculated the ground state of $H_{\text{nc}}$, noted by $|\psi_{\text{nc}}^{\text{g}} \rangle$. For a general molecular Hamiltonian, it is mostly not a noncontextual one. In order to approach the true ground state's energy $E_{\text{g}}$, we need to consider the $H_{\text{c}}$, the contextual component of $H_{\text{qubit}}$. The quantum correction is not necessary to be directly solving $H_{\text{c}}$, 
\begin{align}
E_{\text{g}} = E_{\text{nc}}^{\text{g}} + E_{\text{c}}^g\neq E_{\text{nc}}^{\text{g}} + min\langle H_{\text{c}} \rangle_{}.
\end{align}
Actually, to impose additional pseudosymmetries to $H_{\text{qubit}}$, we need to achieve a unitary rotation such that for some qubit position $i \in Q_{\text{fixed}}$, it only survives the single-qubit Pauli operator $\sigma_{p}^{(i)}$,
$$[UH_{\text{qubit}}U^{^{\dagger}}, \, \sigma_{p}^{(i)}] =0, $$
which is analogous to qubit tapering. When dealing with real, physical symmetry, $U_{r}$ is a Clifford operator and must satisfy 
\begin{align}
U_{r}gU_{r}^{^{\dagger}}=\sigma_{p}^{(i)}, \, \forall g \in G.
\end{align}
The difference in CS method is that the set of generator of CS method and the unitary operation are both being modified, highly correlate to the constructed operator $R(\mathbf{r})$, 
\begin{align}
\tilde{U_{r}}:=U_{\text{C}}U_{\text{G}}, \, U_{\text{C}}R(\mathbf{r})U_{\text{C}}^{\dagger}=\sigma_{p}^{(i)}.
\end{align}
In such case, $U_{\text{C}}$ is no longer a Clifford operation because it maps several Pauli strings into one string. The reason why one can tune the number of qubits in the CS method lies in one can tune the number of generators in the modified noncontextual operator set which unites $R(\mathbf{r})$ with the original one. After the rotation $\tilde{U_{r}}$, one needs to project the Hamiltonian into the corresponding pseudosymmetry sector to actually achieve dimension reduction, i.e., taper qubits by CS method.
By absorbing the calculated $E_{\text{nc}}^{\text{g}}$, one may directly gain the quantum-corrected energy solving the modified contextual Hamiltonian 
\begin{align}
\tilde{H_{\text{c}}}=E_{\text{nc}}^{\text{g}}\mathbf{I} +O_{p}\tilde{U_{r}}H_{\text{c}}\tilde{U_{r}}^{\dagger}O_{p}^{\dagger},
\end{align}
where $\mathbf{I}$ is the identity operator and $O_{p}$ is the projection operator. We refer the reader to ~\cite{weaving2023stabilizer} for complete technical details on contextual subspace construction.

Using the hybrid algorithm CS-VQE to solve the ground-state problem has been demonstrated through numerical and real quantum device experiments. However, the excited-state problems have not yet been verified within the contextual subspace framework.

VQD solves the excited-state problem by tuning the VQE algorithm's cost function, more specifically, by including the overlap term between the two states, 
\begin{align}
  \label{cost_function}
 f_{\text{cost}}&=\langle \psi_{\text{ref}}| U(\theta_{k})^{\dagger}H_{\text{qubit}}U(\theta_{k}) | \psi_{\text{ref}} \rangle + \notag \\
 &\sum_{j=0}^{j<k}\beta_{j}\langle \psi_{\text{ref}}| U(\theta_{j})^{\dagger}U(\theta_{k}) | \psi_{\text{ref}} \rangle \notag \\
 &=\sum_{p} h_{p}\langle \psi_{k} | P | \psi_{k} \rangle + \sum_{j=0}^{j<k} \beta_{j}\langle \psi_{j} | \psi_{k}  \rangle,
\end{align}
where $p$ is the total number of pauli terms in $H_{\text{qubit}}$. Note that $p$ reduces as we use the contextual subspace method to calculate the excited state, leaving only the Pauli terms exist in $\tilde{H_{\text{c}}}$. For instance, if the first excited state aims to be solved, then the overlap term is obtained by the inner product between the calculated ground state and the state after the parameterized unitary circuit under a reference state. Obviously, the excited states are calculated by order using this method. Note that this method is time-consuming in calculating the overlap term when running on a real quantum computer. Overlap-independent excitation techniques like the EOM method~\cite{ollitrault2020quantum} offer computational advantages by avoiding overlap calculations, albeit at the cost of stricter convergence criteria for reference states and reduced robustness against measurement uncertainties. Utilizing the $\tilde{H_{\text{c}}}$ to construct the loss function means that one searches for the excited state in a contextual subspace.

\section{Numerical results}\label{Sec: numericalresults}

\subsection{CS-VQD using problem-inspired ansatz}
\begin{figure}[t]
  \centering
  \includegraphics[width=\linewidth]{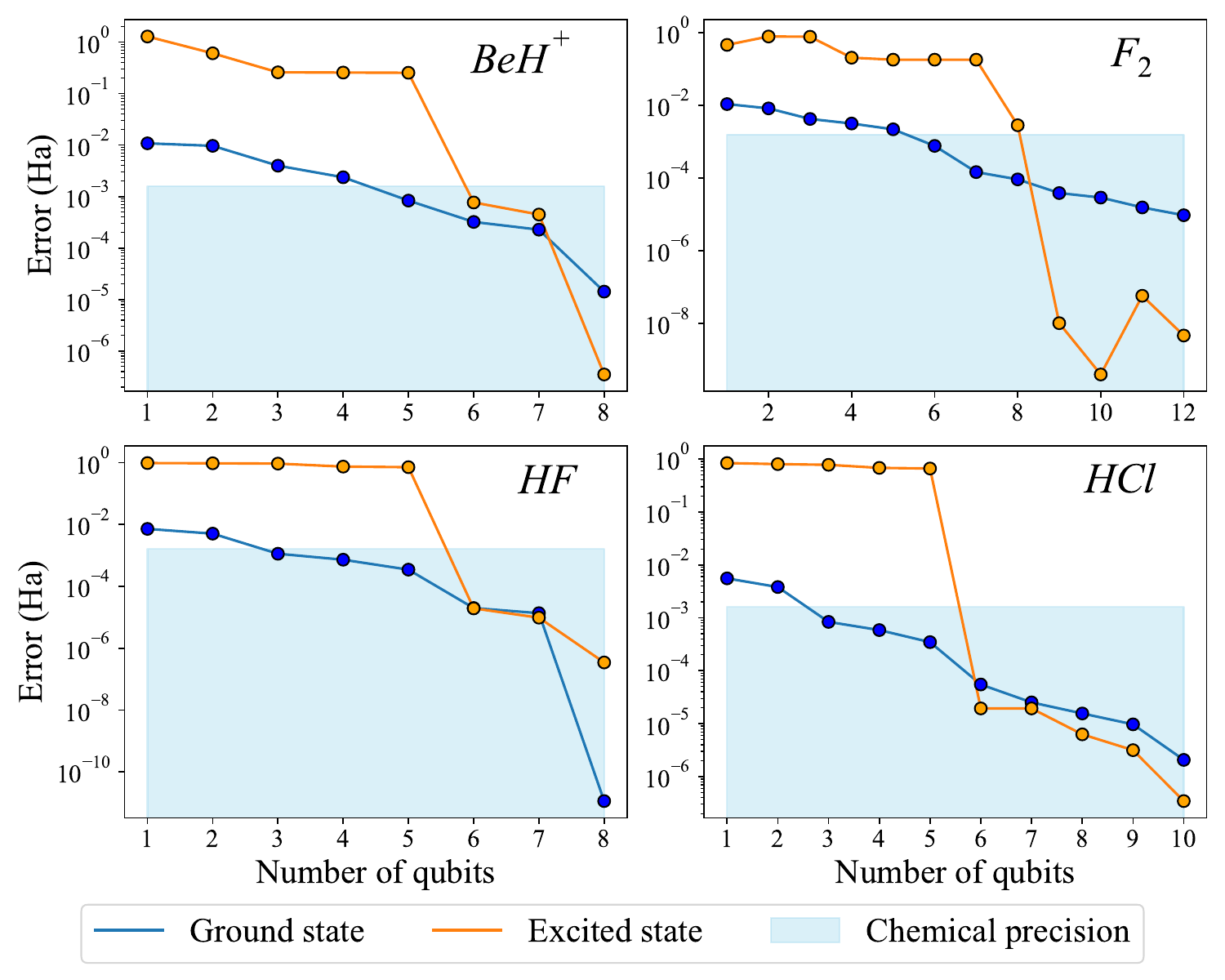}
  \caption{CS-VQE and CS-VQD calculation errors using a UCCSD ansatz versus number of qubits.}
  \label{}
\end{figure}

\begin{figure}[t]
  \centering
  \includegraphics[width=\linewidth]{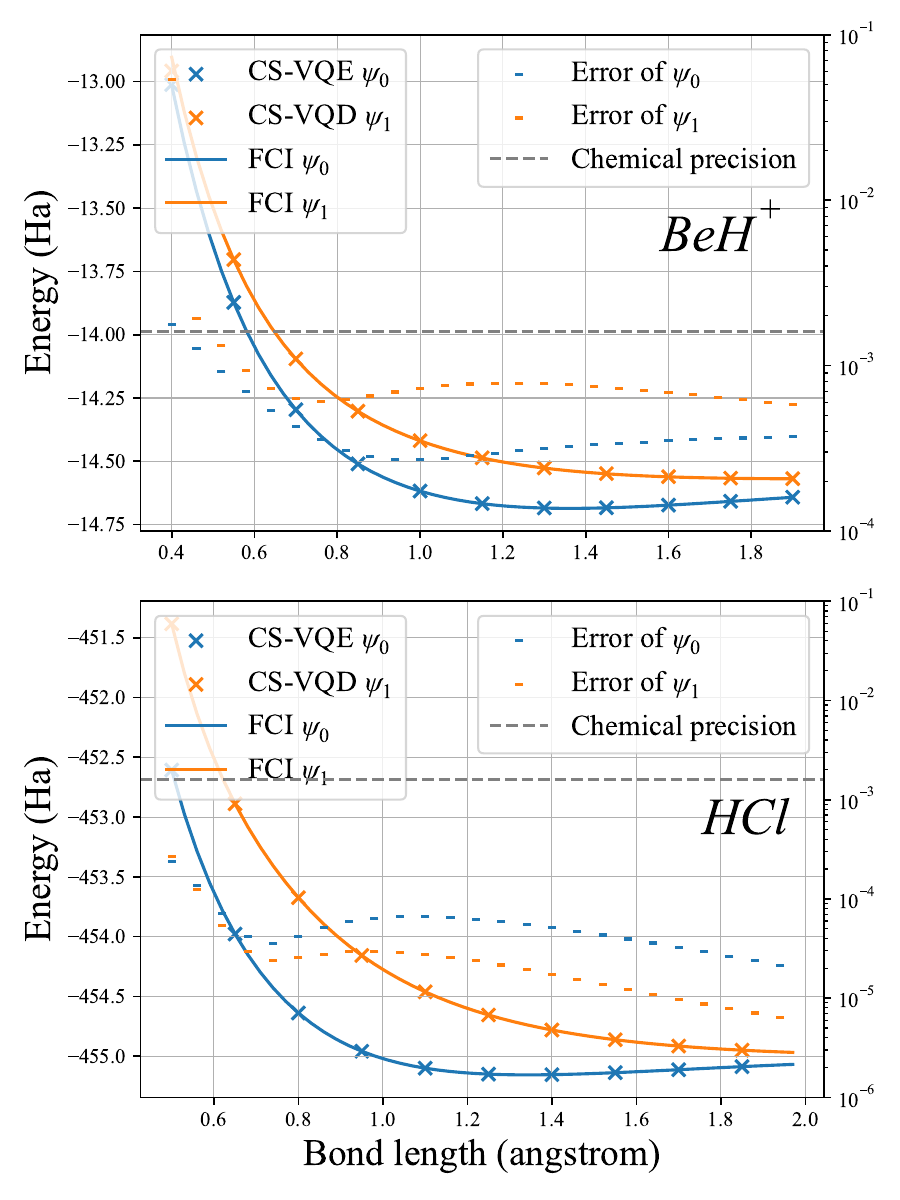}
  \caption{PES of $BeH^+$ and $HCl$ calculated by CS-VQD, using a UCCSD ansatz. $\psi_0$ and $\psi_1$ stand for the ground and the first excited state, respectively. The errors are the absolute value of the difference between FCI and ground or excited states.}
  \label{}
\end{figure}

\begin{figure*}[t]
  \centering
  \includegraphics[width=\linewidth]{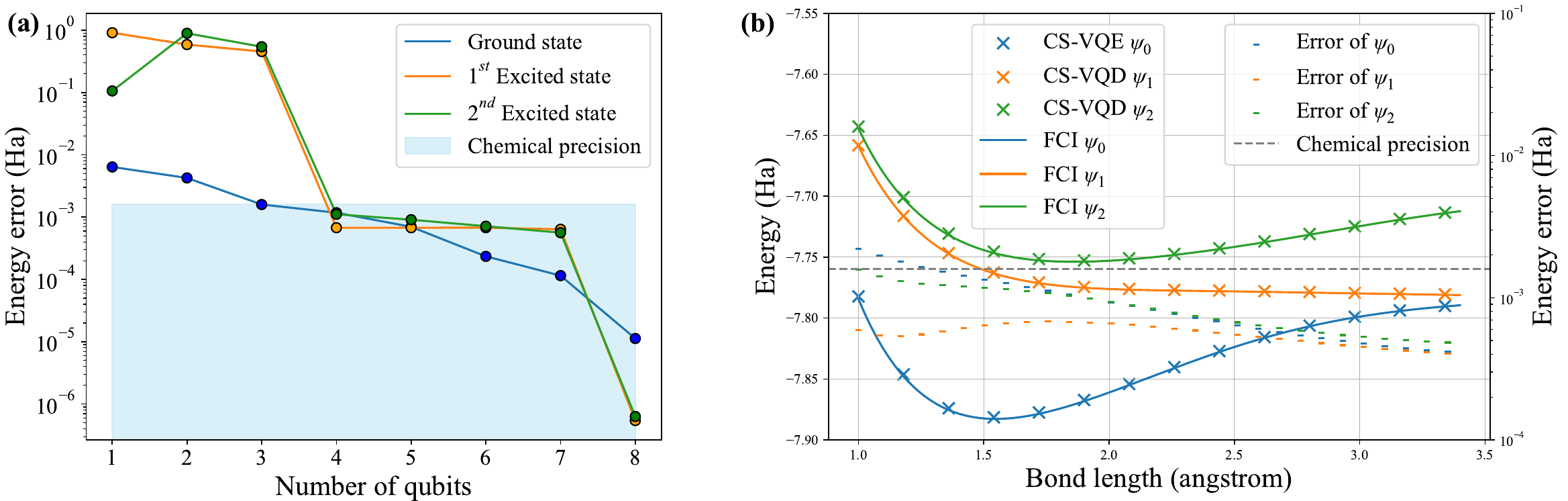}
  \caption{(a)Using different number of qubits to calculate the energy of three states at a fixed bond length $r=1.57473$\AA~between the $Li$ and $H$ atom. (b)Two excited-state PES of $LiH$ calculated by CS-VQD in a 4-qubit usage. } 
  \label{Fig: LiH}
\end{figure*}

We used this CS-VQD method to calculate four molecules using an extension code built on the Symmer~\cite{ralli2022symmer} framework and using Qiskit~\cite{gadi_aleksandrowicz_2019_2562111} as a quantum circuit builder. The Hamiltonian of molecules was constructed using the OpenFermion~\cite{mcclean2020openfermion} program. The full configuration interaction (FCI) method was used to produce the benchmark results on the basis of STO-3G. The complete VQD calculations require 8, 8, 15, and 16 qubits for $BeH^+, \, HF, \, F_2$ and $HCl$, respectively. 
For $F_{2}$ and $HCl$, using qubits more than 10 is really time consuming, therefore we truncate the demonstration at 12 and 10 qubits as long as it shows the performance of CS-VQD. As shown, using CS-VQD can definitely reduce the number of qubits used compared to using VQE solely. We also depict the ground-state results obtained from CS-VQE (CS-VQD decays to CS-VQE when ignoring the overlap term). Compared to ground-state calculation, more qubits are required in order to reach the chemical precision ($1.6 \times 10^{-3}$ hartree) for CS-VQD which is reasonable because the cost function contains more parts and thus leads to a more complex searching route. For promising validity, the ansatz we use here is the widely used Unitary Coupled Cluster Singles and Doubles (UCCSD)~\cite{anand2022quantum, greene-diniz2020generalized}, where the excitation terms can be written as,
\begin{align}
  \label{excitation_operator}
  \hat{T}_{\text{UCCSD}} &= \hat{T}_1 + \hat{T}_2 \notag \\
  &= \sum_{\substack{i \in \text{virt}, \\ \alpha \in \text{occ}}} t_i^\alpha \hat{a}_i^\dagger \hat{a}_\alpha + \sum_{\substack{i,j \in \text{virt}, \\ \alpha,\beta \in \text{occ}}} t_{ij}^{\alpha\beta} \hat{a}_i^\dagger \hat{a}_j^\dagger \hat{a}_\beta \hat{a}_\alpha.
\end{align}
The excitation operators can later be used to construct the unitary circuit in an exponential form,
\begin{equation}
  \label{}
|\psi(\vec{\theta})\rangle = U(\vec{\theta})|\psi_{ref}\rangle = e^{\hat{T}_{\text{UCCSD}} - \hat{T}^\dagger_{\text{UCCSD}}} |\psi_{ref}\rangle.
\end{equation}
The excited-state energy errors of these four molecules all manifest a rapid descent as the number of qubits increases, while it is smoother in the ground-state situation. Based on the full number of qubits used in VQD, we can now reduce the number of qubit in CS-VQD algorithm, which is 2, 2, 6, and 10 qubits for $BeH^+, \, HF, \, F_2$, and $HCl$, respectively.
The code used to generate this paper's numerical results can be found on Github~\cite{yao2025csvqd}.

Using CS-VQD, we also verify the feasibility of calculating other configurations through the excited state's potential energy surfaces of $BeH^+$ and $HCl$. We choose the least number of qubits that achieves the chemical precision for these two molecules, which are both 6. Basically, the landscapes give a precise excited-state energy result. For $HCl$, the obtained excited-state energies even have a higher precision compared to the ground-state energies. This may be attributed to a less significant electronic correlation in this low-order excited state. In addition, if we calculate the landscape by gradually shortening or stretching the bond length as the sample points, we can fit the optimized parameters of the last point into the next point as initial parameters. In this way, we can quickly come to a convergence and obtain a consecutive parameter set and an error curve, which shows a potential to use machine learning techniques to construct a more dense potential energy surface (PES)~\cite{schultz2004efficient}.

To demonstrate the generality of CS-VQD at other molecular configurations, we calculate the PES of LiH using four qubits, including two excited states, see Figure~\ref{Fig: LiH}(b). For the first and second excited states, the energy errors are all less than the chemical accuracy along the curve, ranging from 1.0 to 3.4\AA. We notice the calculated second excited-state's energy error progressively approaches the chemical accuracy, for $LiH$, $BeH^+$, and $HCl$. This corresponds to several reasons, which also imply the different probabilities of success using the CS method to deal with ground and excited states. The excited state's electronic structure changes rapidly as two atom approach, making the solution space more complex and furthermore likely result in a not fully optimized state. From the FCI results, the gap between these two excited states is narrowing, challenging the algorithm to distinguish these two states. Compared to the exact diagonalization method, the VQE-based quantum algorithm is designed to optimize the energy rather than the wave function, although the fidelity between the true electronic state and the quantum calculated state should be small. To gain an improvement in the relatively strong correlation region (the short-length region), one might consider employing a better reference state or adjusting the optimization strategy.

From Figure~\ref{Fig: LiH}(a), we find that the first and second excited states share a similar descent trend when increasing the number of qubits. The energy error also occurs in a rapid drop when the number of qubits is increased from 3 to 4. However, the precision has little increase while increasing the number of qubits from 4 to 7 for both excited states. Besides, we also tried to combine Subspace Search Variational Quantum Eigensolver (SSVQE) with CS but failed in the lack of reaching the chemical precision in most cases based on the same settings mentioned before. This challenge may arise from the inherent difficulty in establishing a reliable reference excited state within the contextual subspace following the projection of the Hamiltonian.

\subsection{Application of spin conservation ansatz}
\begin{figure}[t]
  \centering
  \includegraphics[width=\linewidth]{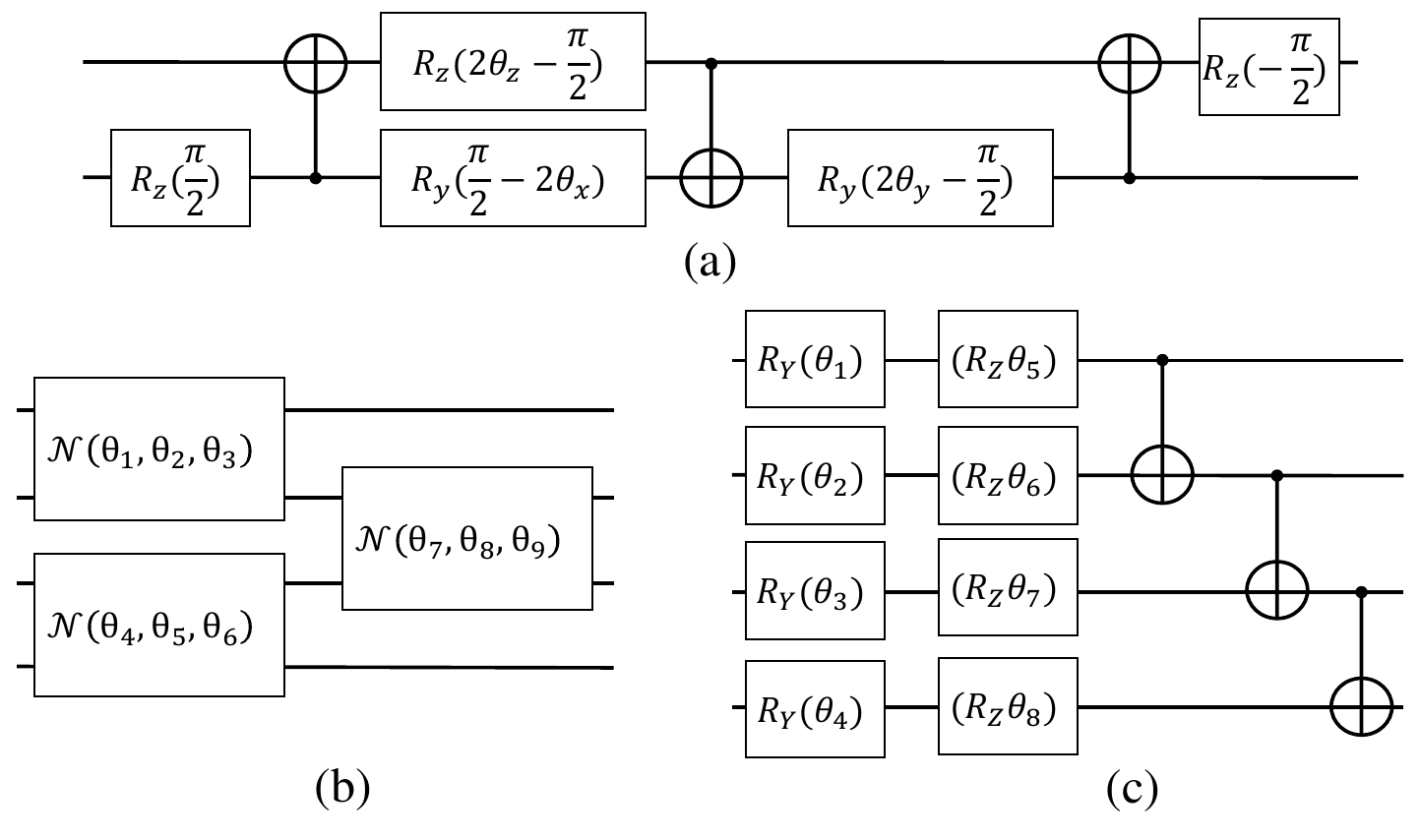}
  \caption{Ansatz Circuit. (a)The realization of $\mathcal{N}(\theta_x,\theta_y,\theta_z)$ block. (b)A single layer of $\mathcal{N}(\theta_x,\theta_y,\theta_z)$ block ansatz circuit in 4-qubit version. (c)A single layer of the $R_{y}R_{z}$ ansatz in 4-qubit version.}
  \label{Fig: ansatz}
\end{figure}

\begin{figure}[t]
  \centering
  \includegraphics[width=\linewidth]{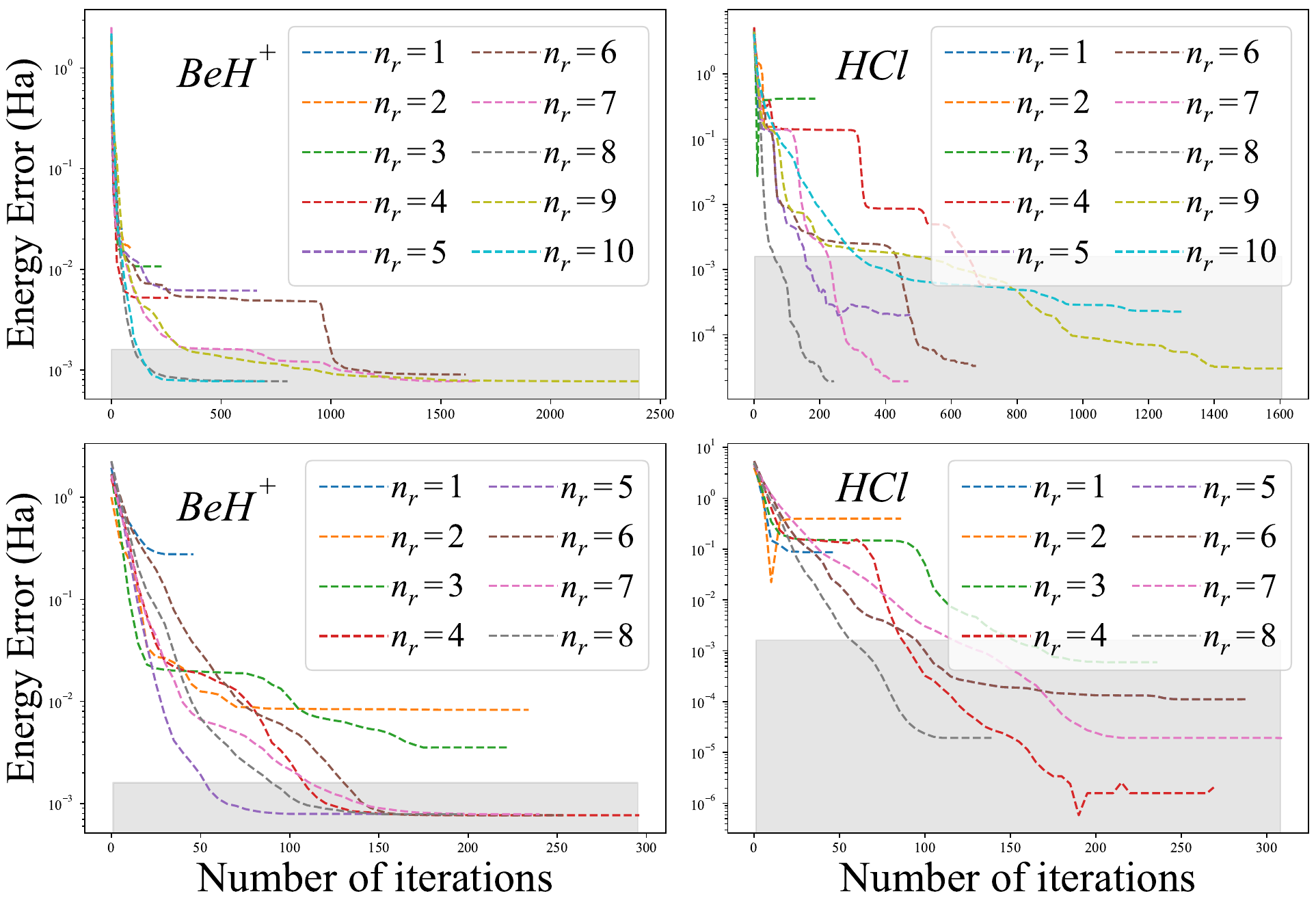}
  \caption{Number of optimization iterations of different repeats. The upper panels show the results using $R_{y}R_{z}$ ansatz while the lower two panels show the results obtained using $\mathcal{N}(\theta_x,\theta_y,\theta_z)$ block ansatz. The grey band in each panel indicates chemical precision. $n_{r}$ means the number of block repetitions in the ansatz.}
  \label{}
\end{figure}

\begin{figure*}[t]
  \includegraphics[width=\linewidth]{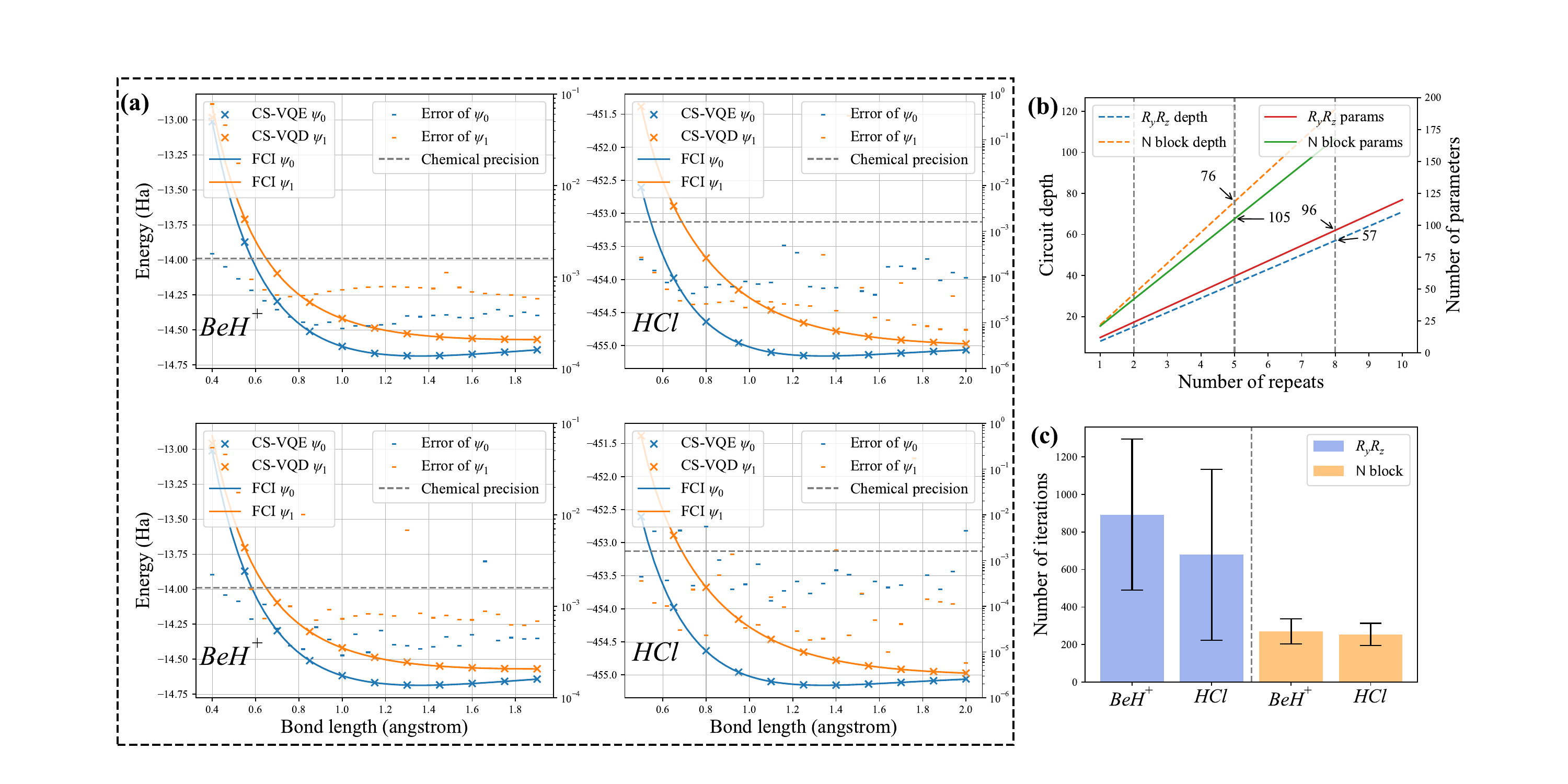}
    \caption{(a) PES of $BeH^{+}$ and $HCl$ calculated by CS-VQD. The upper and lower two panels show the results using $R_{y}R_{z}$ and $\mathcal{N}(\theta_x,\theta_y,\theta_z)$ block ansatz, respectively. In these four panels, the right vertical axis represents the error (in units of hartree).
  (b) In $R_{y}R_{z}$ and $\mathcal{N}(\theta_x,\theta_y,\theta_z)$ ansatz, the number of parameters and circuit depth grow linearly with the number of repeats.
  (c) A significant divergence in iteration counts emerges between these two ansätze.
  }
\end{figure*}

Training the quantum-classical hybrid models is an essential step towards the practical usage of NISQ computers, and an ill-defined ansatz can even be untrainable~\cite{leone2024practical}. The quantum computer tackles the classically hard parts of the training process, which are evolving the circuit and then obtaining an estimation of certain observables. As mentioned before, we used UCCSD~\cite{anand2022quantum, greene-diniz2020generalized}, a chemically inspired one, as the ansatz in computing the excited state. On the other hand, the most famous ansatz is the so-called Hardware Efficient Ansatz (HEA)\cite{kandala2017hardwareefficient} which is designed to mitigate the hardware noise by avoiding non-native gate set usage. Compared to HEA ansätze, UCCSD ansatz might have the ability to avoid Barren Plateaus (BP)~\cite{mcclean2018barren} due to its limited expression ability. However, a recent study~\cite{mao2024determining} shows that a UCC-based ansatz (an alternated disentangled UCC, dUCC) may also suffer from BP problems theoretically. This inspires us to apply HEA, in turn, to verify such an algorithm in a more hardware-efficient way.

Firstly, we adopt one of the most commonly used HEA, the $R_{y}R_{z}$ ansatz~\cite{dcunha2023challenges} , which also shown in Figure~\ref{Fig: ansatz}(c). This ansatz has a simple structure that applies $R_{y}(\theta_{i})$ and $R_{z}(\phi_{i})$ to every qubit and then adds CNOT gates between all qubits. Taking into account the time consumption and actual abilities of the device, we need to determine the optimal repeats of the $R_{y}R_{z}$ ansatz. Thus we test the single-point energy error in different times of repetition using the fixed qubit number $N_{\text{qubit}}=6$ for both $BeH^+$ and $HCl$. 

If one uses a HEA ansatz, it is important to choose a parameter initialization way. Using the aforementioned \textit{continuous} way seems efficient, but it only suits PES or other applications where there already exist correlated data points. Furthermore, when the final optimized parameter configuration converges to an unfavorable state within the parameter landscape, the limited gradient information induced by marginal variations in the loss function substantially constrains the system's ability to escape local minima.
Using a random initialization seems fair; however, it has a non-negligible fluctuation in the final result and may suffer from BP problems. Setting all the parameters to zero seems more friendly to the BP problem but needs more evidence to prove.
For the single-point test, we take the parameter initialization in a random manner, with bond length $r=1.3447$\AA~and $r=1.3413$\AA~for $BeH^+$ and $HCl$, respectively. 

Contextual subspace method can be seen as imposing pseudosymmetries in a way. However, after the original Hamiltonian is projected into the contextual subspace, we wonder if there is any real symmetry left in the system. Using the variational quantum algorithm to solve a molecular problem inherently maintains two aspects to consider the whole system's symmetry, one from the quantum circuit and the other from the molecular itself. Here, we consider a property from the molecular, the spin symmetry.
We adopt the so-called $\mathcal{N}(\theta_x,\theta_y,\theta_z)$ block~\cite{lyu2023symmetry, vatan2004optimal} ansatz, a symmetry enhanced circuit to verify if we can use the conservation of $S_{z}$ to help with the optimization procedure. This entangling block gate contains a more complex form of entanglement as shown in Figure~\ref{Fig: ansatz}(a).

At these specific bond lengths, the less repeat block achieving chemical precision is 6 and 4 for $R_{y}R_{z}$ ansatz and 5 and 4 for $\mathcal{N}(\theta_x,\theta_y,\theta_z)$ block ansatz. Notably, if one uses a zero parameter initialization rather than the random way, the smaller number of repeat blocks does not mean any block number that is larger than that would not certainly satisfy the chemical precision. 

To ensure that at every other bond length, the used ansatz would also work well, we adopt the use of 8 and 5 repeat blocks for the $R_{y}R_{z}$ ansatz and the $\mathcal{N}(\theta_x,\theta_y,\theta_z)$ block ansatz, respectively. Also, in this repeat block setting, we can have a similar circuit depth and number of parameters, which can help us to compare the performance between these two circuits in a more reasonable way.
Comparing the PES calculated using the $R_{y}R_{z}$ and $\mathcal{N}(\theta_x,\theta_y,\theta_z)$ block ansatz, we find that both ansätze can help to achieve the chemical precision.
While the $R_{y}R_{z}$ ansatz shows a more stable performance in the bond length range, the $\mathcal{N}(\theta_x,\theta_y,\theta_z)$ block ansatz shows an obvious advantage in iteration number. 
Here we statistically analyzed the times of optimization iteration on the sampled points of the PES. With a slightly deeper circuit depth, the $\mathcal{N}(\theta_x,\theta_y,\theta_z)$ block ansatz needs less times of iteration on an average estimate. To $BeH^+$ and $HCl$, using the $R_{y}R_{z}$ ansatz needs 892.6 and 678.3 times of iteration while the $\mathcal{N}(\theta_x,\theta_y,\theta_z)$ block ansatz only needs 270.5 and 254.2 times of iteration.
Also, this symmetry-utilized ansatz has a smaller deviation, showing more stable performance. More specifically, with 402.5 and 454.6 times using $R_{y}R_{z}$ ansatz compared to 67 and 59 times using $\mathcal{N}(\theta_x,\theta_y,\theta_z)$ block ansatz, for $BeH^+$ and $HCl$, respectively.

\section{Conclusion}
In this work, we have proposed and demonstrated a resource-efficient framework for excited-state calculations in quantum chemistry by integrating the contextual subspace method with the VQD algorithm, referred to as CS-VQD. This approach significantly reduces the qubit requirements for excited-state calculations on NISQ hardware. Our numerical results demonstrate that CS-VQD achieves chemical precision for molecules like $BeH^+$, $HF$, $F2$, and $HCl$. We also showed that the spin-conserving $\mathcal{N}(\theta_x,\theta_y,\theta_z)$ansatz significantly reduces optimization iterations compared to the $R_{y}R_{z}$ ansatz. Future work could explore the application of machine learning techniques to further optimize the parameter sets and improve the convergence of these algorithms, especially in regions of strong electronic correlation. Although combining CS with SSVQE proved ineffective, our findings highlight the potential of CS-VQD for efficient excited-state calculations, paving the way for practical quantum chemistry applications on near-term quantum devices.

\begin{acknowledgments}
This work is supported in part by the National Natural Science Foundation of China under Grant 62304037, in part by the Natural Science Foundation of Jiangsu Province under Grant BK20230828, in part by the Jiangsu Provincial Scientific Research Center of Applied Mathematics under Grant BK20233002, in part by the Young Elite Scientists Sponsorship Program by CAST under Grant 2022QNRC001, in part by the Southeast University Interdisciplinary Research Program for Young Scholars under Grant 2024FGC1005, in part by the Fundamental Research Funds for the Central Universities under Grant 2242024K40016, and in part by the Start-up Research Fund of Southeast University under Grant RF1028623173.
\end{acknowledgments}

\bibliography{csvqd_main}

\begin{thebibliography}{35}%
\makeatletter
\providecommand \@ifxundefined [1]{%
 \@ifx{#1\undefined}
}%
\providecommand \@ifnum [1]{%
 \ifnum #1\expandafter \@firstoftwo
 \else \expandafter \@secondoftwo
 \fi
}%
\providecommand \@ifx [1]{%
 \ifx #1\expandafter \@firstoftwo
 \else \expandafter \@secondoftwo
 \fi
}%
\providecommand \natexlab [1]{#1}%
\providecommand \enquote  [1]{``#1''}%
\providecommand \bibnamefont  [1]{#1}%
\providecommand \bibfnamefont [1]{#1}%
\providecommand \citenamefont [1]{#1}%
\providecommand \href@noop [0]{\@secondoftwo}%
\providecommand \href [0]{\begingroup \@sanitize@url \@href}%
\providecommand \@href[1]{\@@startlink{#1}\@@href}%
\providecommand \@@href[1]{\endgroup#1\@@endlink}%
\providecommand \@sanitize@url [0]{\catcode `\\12\catcode `\$12\catcode `\&12\catcode `\#12\catcode `\^12\catcode `\_12\catcode `\%12\relax}%
\providecommand \@@startlink[1]{}%
\providecommand \@@endlink[0]{}%
\providecommand \url  [0]{\begingroup\@sanitize@url \@url }%
\providecommand \@url [1]{\endgroup\@href {#1}{\urlprefix }}%
\providecommand \urlprefix  [0]{URL }%
\providecommand \Eprint [0]{\href }%
\providecommand \doibase [0]{https://doi.org/}%
\providecommand \selectlanguage [0]{\@gobble}%
\providecommand \bibinfo  [0]{\@secondoftwo}%
\providecommand \bibfield  [0]{\@secondoftwo}%
\providecommand \translation [1]{[#1]}%
\providecommand \BibitemOpen [0]{}%
\providecommand \bibitemStop [0]{}%
\providecommand \bibitemNoStop [0]{.\EOS\space}%
\providecommand \EOS [0]{\spacefactor3000\relax}%
\providecommand \BibitemShut  [1]{\csname bibitem#1\endcsname}%
\let\auto@bib@innerbib\@empty
\bibitem [{\citenamefont {Preskill}(2018)}]{preskill2018quantum}%
  \BibitemOpen
  \bibfield  {author} {\bibinfo {author} {\bibfnamefont {J.}~\bibnamefont {Preskill}},\ }\bibfield  {title} {\bibinfo {title} {Quantum {{Computing}} in the {{NISQ}} era and beyond},\ }\href {https://doi.org/10.22331/q-2018-08-06-79} {\bibfield  {journal} {\bibinfo  {journal} {Quantum}\ }\textbf {\bibinfo {volume} {2}},\ \bibinfo {pages} {79} (\bibinfo {year} {2018})}\BibitemShut {NoStop}%
\bibitem [{\citenamefont {Kitaev}(1995)}]{kitaev1995quantum}%
  \BibitemOpen
  \bibfield  {author} {\bibinfo {author} {\bibfnamefont {A.~Y.}\ \bibnamefont {Kitaev}},\ }\href {http://arxiv.org/abs/quant-ph/9511026} {\bibinfo {title} {Quantum measurements and the {{Abelian Stabilizer Problem}}}} (\bibinfo {year} {1995})\BibitemShut {NoStop}%
\bibitem [{\citenamefont {{Google Quantum AI and Collaborators}}\ \emph {et~al.}(2024)\citenamefont {{Google Quantum AI and Collaborators}} \emph {et~al.}}]{googlequantumaiandcollaborators2024quantum}%
  \BibitemOpen
  \bibfield  {author} {\bibinfo {author} {\bibnamefont {{Google Quantum AI and Collaborators}}} \emph {et~al.},\ }\bibfield  {title} {\bibinfo {title} {Quantum error correction below the surface code threshold},\ }\bibfield  {journal} {\bibinfo  {journal} {Nature}\ }\href {https://doi.org/10.1038/s41586-024-08449-y} {10.1038/s41586-024-08449-y} (\bibinfo {year} {2024})\BibitemShut {NoStop}%
\bibitem [{\citenamefont {Caesura}\ \emph {et~al.}(2025)\citenamefont {Caesura}, \citenamefont {Cortes}, \citenamefont {Pol}, \citenamefont {Sim}, \citenamefont {Steudtner}, \citenamefont {Anselmetti}, \citenamefont {Degroote}, \citenamefont {Moll}, \citenamefont {Santagati}, \citenamefont {Streif},\ and\ \citenamefont {Tautermann}}]{caesura2025faster}%
  \BibitemOpen
  \bibfield  {author} {\bibinfo {author} {\bibfnamefont {A.}~\bibnamefont {Caesura}}, \bibinfo {author} {\bibfnamefont {C.~L.}\ \bibnamefont {Cortes}}, \bibinfo {author} {\bibfnamefont {W.}~\bibnamefont {Pol}}, \bibinfo {author} {\bibfnamefont {S.}~\bibnamefont {Sim}}, \bibinfo {author} {\bibfnamefont {M.}~\bibnamefont {Steudtner}}, \bibinfo {author} {\bibfnamefont {G.-L.~R.}\ \bibnamefont {Anselmetti}}, \bibinfo {author} {\bibfnamefont {M.}~\bibnamefont {Degroote}}, \bibinfo {author} {\bibfnamefont {N.}~\bibnamefont {Moll}}, \bibinfo {author} {\bibfnamefont {R.}~\bibnamefont {Santagati}}, \bibinfo {author} {\bibfnamefont {M.}~\bibnamefont {Streif}},\ and\ \bibinfo {author} {\bibfnamefont {C.~S.}\ \bibnamefont {Tautermann}},\ }\href {http://arxiv.org/abs/2501.06165} {\bibinfo {title} {Faster quantum chemistry simulations on a quantum computer with improved tensor factorization and active volume compilation}} (\bibinfo {year} {2025})\BibitemShut {NoStop}%
\bibitem [{\citenamefont {Huang}\ \emph {et~al.}(2022)\citenamefont {Huang}, \citenamefont {Govoni},\ and\ \citenamefont {Galli}}]{huang2022simulating}%
  \BibitemOpen
  \bibfield  {author} {\bibinfo {author} {\bibfnamefont {B.}~\bibnamefont {Huang}}, \bibinfo {author} {\bibfnamefont {M.}~\bibnamefont {Govoni}},\ and\ \bibinfo {author} {\bibfnamefont {G.}~\bibnamefont {Galli}},\ }\bibfield  {title} {\bibinfo {title} {Simulating the {{Electronic Structure}} of {{Spin Defects}} on {{Quantum Computers}}},\ }\href {https://doi.org/10.1103/PRXQuantum.3.010339} {\bibfield  {journal} {\bibinfo  {journal} {PRX Quantum}\ }\textbf {\bibinfo {volume} {3}},\ \bibinfo {pages} {010339} (\bibinfo {year} {2022})}\BibitemShut {NoStop}%
\bibitem [{\citenamefont {Ahmad}\ \emph {et~al.}(2016)\citenamefont {Ahmad}, \citenamefont {Ahmed}, \citenamefont {Anwar}, \citenamefont {Sheraz},\ and\ \citenamefont {Sikorski}}]{ahmad2016photostability}%
  \BibitemOpen
  \bibfield  {author} {\bibinfo {author} {\bibfnamefont {I.}~\bibnamefont {Ahmad}}, \bibinfo {author} {\bibfnamefont {S.}~\bibnamefont {Ahmed}}, \bibinfo {author} {\bibfnamefont {Z.}~\bibnamefont {Anwar}}, \bibinfo {author} {\bibfnamefont {M.~A.}\ \bibnamefont {Sheraz}},\ and\ \bibinfo {author} {\bibfnamefont {M.}~\bibnamefont {Sikorski}},\ }\bibfield  {title} {\bibinfo {title} {Photostability and {{Photostabilization}} of {{Drugs}} and {{Drug Products}}},\ }\href {https://doi.org/10.1155/2016/8135608} {\bibfield  {journal} {\bibinfo  {journal} {International Journal of Photoenergy}\ }\textbf {\bibinfo {volume} {2016}},\ \bibinfo {pages} {1} (\bibinfo {year} {2016})}\BibitemShut {NoStop}%
\bibitem [{\citenamefont {Fedorov}\ and\ \citenamefont {Gelfand}(2021)}]{fedorov2021practical}%
  \BibitemOpen
  \bibfield  {author} {\bibinfo {author} {\bibfnamefont {A.~K.}\ \bibnamefont {Fedorov}}\ and\ \bibinfo {author} {\bibfnamefont {M.~S.}\ \bibnamefont {Gelfand}},\ }\bibfield  {title} {\bibinfo {title} {Towards practical applications in quantum computational biology},\ }\href {https://doi.org/10.1038/s43588-021-00024-z} {\bibfield  {journal} {\bibinfo  {journal} {Nature Computational Science}\ }\textbf {\bibinfo {volume} {1}},\ \bibinfo {pages} {114} (\bibinfo {year} {2021})}\BibitemShut {NoStop}%
\bibitem [{\citenamefont {Beard}\ \emph {et~al.}(2019)\citenamefont {Beard}, \citenamefont {Sivaraman}, \citenamefont {{V{\'a}zquez-Mayagoitia}}, \citenamefont {Vishwanath},\ and\ \citenamefont {Cole}}]{beard2019comparative}%
  \BibitemOpen
  \bibfield  {author} {\bibinfo {author} {\bibfnamefont {E.~J.}\ \bibnamefont {Beard}}, \bibinfo {author} {\bibfnamefont {G.}~\bibnamefont {Sivaraman}}, \bibinfo {author} {\bibfnamefont {{\'A}.}~\bibnamefont {{V{\'a}zquez-Mayagoitia}}}, \bibinfo {author} {\bibfnamefont {V.}~\bibnamefont {Vishwanath}},\ and\ \bibinfo {author} {\bibfnamefont {J.~M.}\ \bibnamefont {Cole}},\ }\bibfield  {title} {\bibinfo {title} {Comparative dataset of experimental and computational attributes of {{UV}}/vis absorption spectra},\ }\href {https://doi.org/10.1038/s41597-019-0306-0} {\bibfield  {journal} {\bibinfo  {journal} {Scientific Data}\ }\textbf {\bibinfo {volume} {6}},\ \bibinfo {pages} {307} (\bibinfo {year} {2019})}\BibitemShut {NoStop}%
\bibitem [{\citenamefont {Bravyi}\ \emph {et~al.}(2017)\citenamefont {Bravyi}, \citenamefont {Gambetta}, \citenamefont {Mezzacapo},\ and\ \citenamefont {Temme}}]{bravyi2017tapering}%
  \BibitemOpen
  \bibfield  {author} {\bibinfo {author} {\bibfnamefont {S.}~\bibnamefont {Bravyi}}, \bibinfo {author} {\bibfnamefont {J.~M.}\ \bibnamefont {Gambetta}}, \bibinfo {author} {\bibfnamefont {A.}~\bibnamefont {Mezzacapo}},\ and\ \bibinfo {author} {\bibfnamefont {K.}~\bibnamefont {Temme}},\ }\href {http://arxiv.org/abs/1701.08213} {\bibinfo {title} {Tapering off qubits to simulate fermionic {{Hamiltonians}}}} (\bibinfo {year} {2017})\BibitemShut {NoStop}%
\bibitem [{\citenamefont {Setia}\ \emph {et~al.}(2020)\citenamefont {Setia}, \citenamefont {Chen}, \citenamefont {Rice}, \citenamefont {Mezzacapo}, \citenamefont {Pistoia},\ and\ \citenamefont {Whitfield}}]{setia2020reducing}%
  \BibitemOpen
  \bibfield  {author} {\bibinfo {author} {\bibfnamefont {K.}~\bibnamefont {Setia}}, \bibinfo {author} {\bibfnamefont {R.}~\bibnamefont {Chen}}, \bibinfo {author} {\bibfnamefont {J.~E.}\ \bibnamefont {Rice}}, \bibinfo {author} {\bibfnamefont {A.}~\bibnamefont {Mezzacapo}}, \bibinfo {author} {\bibfnamefont {M.}~\bibnamefont {Pistoia}},\ and\ \bibinfo {author} {\bibfnamefont {J.~D.}\ \bibnamefont {Whitfield}},\ }\bibfield  {title} {\bibinfo {title} {Reducing {{Qubit Requirements}} for {{Quantum Simulations Using Molecular Point Group Symmetries}}},\ }\href {https://doi.org/10.1021/acs.jctc.0c00113} {\bibfield  {journal} {\bibinfo  {journal} {Journal of Chemical Theory and Computation}\ }\textbf {\bibinfo {volume} {16}},\ \bibinfo {pages} {6091} (\bibinfo {year} {2020})}\BibitemShut {NoStop}%
\bibitem [{\citenamefont {Kirby}\ and\ \citenamefont {Love}(2019)}]{kirby2019contextuality}%
  \BibitemOpen
  \bibfield  {author} {\bibinfo {author} {\bibfnamefont {W.~M.}\ \bibnamefont {Kirby}}\ and\ \bibinfo {author} {\bibfnamefont {P.~J.}\ \bibnamefont {Love}},\ }\bibfield  {title} {\bibinfo {title} {Contextuality {{Test}} of the {{Nonclassicality}} of {{Variational Quantum Eigensolvers}}},\ }\href {https://doi.org/10.1103/PhysRevLett.123.200501} {\bibfield  {journal} {\bibinfo  {journal} {Physical Review Letters}\ }\textbf {\bibinfo {volume} {123}},\ \bibinfo {pages} {200501} (\bibinfo {year} {2019})}\BibitemShut {NoStop}%
\bibitem [{\citenamefont {Kirby}\ \emph {et~al.}(2021)\citenamefont {Kirby}, \citenamefont {Tranter},\ and\ \citenamefont {Love}}]{kirby2021contextual}%
  \BibitemOpen
  \bibfield  {author} {\bibinfo {author} {\bibfnamefont {W.~M.}\ \bibnamefont {Kirby}}, \bibinfo {author} {\bibfnamefont {A.}~\bibnamefont {Tranter}},\ and\ \bibinfo {author} {\bibfnamefont {P.~J.}\ \bibnamefont {Love}},\ }\bibfield  {title} {\bibinfo {title} {Contextual {{Subspace Variational Quantum Eigensolver}}},\ }\href {https://doi.org/10.22331/q-2021-05-14-456} {\bibfield  {journal} {\bibinfo  {journal} {Quantum}\ }\textbf {\bibinfo {volume} {5}},\ \bibinfo {pages} {456} (\bibinfo {year} {2021})}\BibitemShut {NoStop}%
\bibitem [{\citenamefont {Higgott}\ \emph {et~al.}(2019)\citenamefont {Higgott}, \citenamefont {Wang},\ and\ \citenamefont {Brierley}}]{higgott2019variational}%
  \BibitemOpen
  \bibfield  {author} {\bibinfo {author} {\bibfnamefont {O.}~\bibnamefont {Higgott}}, \bibinfo {author} {\bibfnamefont {D.}~\bibnamefont {Wang}},\ and\ \bibinfo {author} {\bibfnamefont {S.}~\bibnamefont {Brierley}},\ }\bibfield  {title} {\bibinfo {title} {Variational {{Quantum Computation}} of {{Excited States}}},\ }\href {https://doi.org/10.22331/q-2019-07-01-156} {\bibfield  {journal} {\bibinfo  {journal} {Quantum}\ }\textbf {\bibinfo {volume} {3}},\ \bibinfo {pages} {156} (\bibinfo {year} {2019})}\BibitemShut {NoStop}%
\bibitem [{\citenamefont {Fedorov}\ \emph {et~al.}(2022)\citenamefont {Fedorov}, \citenamefont {Peng}, \citenamefont {Govind},\ and\ \citenamefont {Alexeev}}]{fedorov2022vqe}%
  \BibitemOpen
  \bibfield  {author} {\bibinfo {author} {\bibfnamefont {D.~A.}\ \bibnamefont {Fedorov}}, \bibinfo {author} {\bibfnamefont {B.}~\bibnamefont {Peng}}, \bibinfo {author} {\bibfnamefont {N.}~\bibnamefont {Govind}},\ and\ \bibinfo {author} {\bibfnamefont {Y.}~\bibnamefont {Alexeev}},\ }\bibfield  {title} {\bibinfo {title} {{{VQE}} method: A short survey and recent developments},\ }\href {https://doi.org/10.1186/s41313-021-00032-6} {\bibfield  {journal} {\bibinfo  {journal} {Materials Theory}\ }\textbf {\bibinfo {volume} {6}},\ \bibinfo {pages} {2} (\bibinfo {year} {2022})}\BibitemShut {NoStop}%
\bibitem [{\citenamefont {Peruzzo}\ \emph {et~al.}(2014)\citenamefont {Peruzzo}, \citenamefont {McClean}, \citenamefont {Shadbolt}, \citenamefont {Yung}, \citenamefont {Zhou}, \citenamefont {Love}, \citenamefont {{Aspuru-Guzik}},\ and\ \citenamefont {O'Brien}}]{peruzzo2014variational}%
  \BibitemOpen
  \bibfield  {author} {\bibinfo {author} {\bibfnamefont {A.}~\bibnamefont {Peruzzo}}, \bibinfo {author} {\bibfnamefont {J.}~\bibnamefont {McClean}}, \bibinfo {author} {\bibfnamefont {P.}~\bibnamefont {Shadbolt}}, \bibinfo {author} {\bibfnamefont {M.-H.}\ \bibnamefont {Yung}}, \bibinfo {author} {\bibfnamefont {X.-Q.}\ \bibnamefont {Zhou}}, \bibinfo {author} {\bibfnamefont {P.~J.}\ \bibnamefont {Love}}, \bibinfo {author} {\bibfnamefont {A.}~\bibnamefont {{Aspuru-Guzik}}},\ and\ \bibinfo {author} {\bibfnamefont {J.~L.}\ \bibnamefont {O'Brien}},\ }\bibfield  {title} {\bibinfo {title} {A variational eigenvalue solver on a photonic quantum processor},\ }\href {https://doi.org/10.1038/ncomms5213} {\bibfield  {journal} {\bibinfo  {journal} {Nature Communications}\ }\textbf {\bibinfo {volume} {5}},\ \bibinfo {pages} {4213} (\bibinfo {year} {2014})}\BibitemShut {NoStop}%
\bibitem [{\citenamefont {Kandala}\ \emph {et~al.}(2017)\citenamefont {Kandala}, \citenamefont {Mezzacapo}, \citenamefont {Temme}, \citenamefont {Takita}, \citenamefont {Brink}, \citenamefont {Chow},\ and\ \citenamefont {Gambetta}}]{kandala2017hardwareefficient}%
  \BibitemOpen
  \bibfield  {author} {\bibinfo {author} {\bibfnamefont {A.}~\bibnamefont {Kandala}}, \bibinfo {author} {\bibfnamefont {A.}~\bibnamefont {Mezzacapo}}, \bibinfo {author} {\bibfnamefont {K.}~\bibnamefont {Temme}}, \bibinfo {author} {\bibfnamefont {M.}~\bibnamefont {Takita}}, \bibinfo {author} {\bibfnamefont {M.}~\bibnamefont {Brink}}, \bibinfo {author} {\bibfnamefont {J.~M.}\ \bibnamefont {Chow}},\ and\ \bibinfo {author} {\bibfnamefont {J.~M.}\ \bibnamefont {Gambetta}},\ }\bibfield  {title} {\bibinfo {title} {Hardware-efficient variational quantum eigensolver for small molecules and quantum magnets},\ }\href {https://doi.org/10.1038/nature23879} {\bibfield  {journal} {\bibinfo  {journal} {Nature}\ }\textbf {\bibinfo {volume} {549}},\ \bibinfo {pages} {242} (\bibinfo {year} {2017})}\BibitemShut {NoStop}%
\bibitem [{\citenamefont {D'Cunha}\ \emph {et~al.}(2023)\citenamefont {D'Cunha}, \citenamefont {Crawford}, \citenamefont {Motta},\ and\ \citenamefont {Rice}}]{dcunha2023challenges}%
  \BibitemOpen
  \bibfield  {author} {\bibinfo {author} {\bibfnamefont {R.}~\bibnamefont {D'Cunha}}, \bibinfo {author} {\bibfnamefont {T.~D.}\ \bibnamefont {Crawford}}, \bibinfo {author} {\bibfnamefont {M.}~\bibnamefont {Motta}},\ and\ \bibinfo {author} {\bibfnamefont {J.~E.}\ \bibnamefont {Rice}},\ }\bibfield  {title} {\bibinfo {title} {Challenges in the {{Use}} of {{Quantum Computing Hardware-Efficient Ans{\"a}tze}} in {{Electronic Structure Theory}}},\ }\href {https://doi.org/10.1021/acs.jpca.2c08430} {\bibfield  {journal} {\bibinfo  {journal} {The Journal of Physical Chemistry A}\ }\textbf {\bibinfo {volume} {127}},\ \bibinfo {pages} {3437} (\bibinfo {year} {2023})}\BibitemShut {NoStop}%
\bibitem [{\citenamefont {Weaving}\ \emph {et~al.}(2023{\natexlab{a}})\citenamefont {Weaving}, \citenamefont {Ralli}, \citenamefont {Love}, \citenamefont {Succi},\ and\ \citenamefont {Coveney}}]{weaving2023contextual}%
  \BibitemOpen
  \bibfield  {author} {\bibinfo {author} {\bibfnamefont {T.}~\bibnamefont {Weaving}}, \bibinfo {author} {\bibfnamefont {A.}~\bibnamefont {Ralli}}, \bibinfo {author} {\bibfnamefont {P.~J.}\ \bibnamefont {Love}}, \bibinfo {author} {\bibfnamefont {S.}~\bibnamefont {Succi}},\ and\ \bibinfo {author} {\bibfnamefont {P.~V.}\ \bibnamefont {Coveney}},\ }\href {http://arxiv.org/abs/2312.04392} {\bibinfo {title} {Contextual {{Subspace Variational Quantum Eigensolver Calculation}} of the {{Dissociation Curve}} of {{Molecular Nitrogen}} on a {{Superconducting Quantum Computer}}}} (\bibinfo {year} {2023}{\natexlab{a}})\BibitemShut {NoStop}%
\bibitem [{\citenamefont {Ralli}\ \emph {et~al.}(2023)\citenamefont {Ralli}, \citenamefont {Weaving}, \citenamefont {Tranter}, \citenamefont {Kirby}, \citenamefont {Love},\ and\ \citenamefont {Coveney}}]{ralli2023unitary}%
  \BibitemOpen
  \bibfield  {author} {\bibinfo {author} {\bibfnamefont {A.}~\bibnamefont {Ralli}}, \bibinfo {author} {\bibfnamefont {T.}~\bibnamefont {Weaving}}, \bibinfo {author} {\bibfnamefont {A.}~\bibnamefont {Tranter}}, \bibinfo {author} {\bibfnamefont {W.~M.}\ \bibnamefont {Kirby}}, \bibinfo {author} {\bibfnamefont {P.~J.}\ \bibnamefont {Love}},\ and\ \bibinfo {author} {\bibfnamefont {P.~V.}\ \bibnamefont {Coveney}},\ }\bibfield  {title} {\bibinfo {title} {Unitary {{Partitioning}} and the {{Contextual Subspace Variational Quantum Eigensolver}}},\ }\href {https://doi.org/10.1103/PhysRevResearch.5.013095} {\bibfield  {journal} {\bibinfo  {journal} {Physical Review Research}\ }\textbf {\bibinfo {volume} {5}},\ \bibinfo {pages} {013095} (\bibinfo {year} {2023})}\BibitemShut {NoStop}%
\bibitem [{\citenamefont {Weaving}\ \emph {et~al.}(2023{\natexlab{b}})\citenamefont {Weaving}, \citenamefont {Ralli}, \citenamefont {Kirby}, \citenamefont {Tranter}, \citenamefont {Love},\ and\ \citenamefont {Coveney}}]{weaving2023stabilizer}%
  \BibitemOpen
  \bibfield  {author} {\bibinfo {author} {\bibfnamefont {T.}~\bibnamefont {Weaving}}, \bibinfo {author} {\bibfnamefont {A.}~\bibnamefont {Ralli}}, \bibinfo {author} {\bibfnamefont {W.~M.}\ \bibnamefont {Kirby}}, \bibinfo {author} {\bibfnamefont {A.}~\bibnamefont {Tranter}}, \bibinfo {author} {\bibfnamefont {P.~J.}\ \bibnamefont {Love}},\ and\ \bibinfo {author} {\bibfnamefont {P.~V.}\ \bibnamefont {Coveney}},\ }\bibfield  {title} {\bibinfo {title} {A {{Stabilizer Framework}} for the {{Contextual Subspace Variational Quantum Eigensolver}} and the {{Noncontextual Projection Ansatz}}},\ }\href {https://doi.org/10.1021/acs.jctc.2c00910} {\bibfield  {journal} {\bibinfo  {journal} {Journal of Chemical Theory and Computation}\ }\textbf {\bibinfo {volume} {19}},\ \bibinfo {pages} {808} (\bibinfo {year} {2023}{\natexlab{b}})}\BibitemShut {NoStop}%
\bibitem [{\citenamefont {Bell}(1964)}]{bell1964einstein}%
  \BibitemOpen
  \bibfield  {author} {\bibinfo {author} {\bibfnamefont {J.~S.}\ \bibnamefont {Bell}},\ }\bibfield  {title} {\bibinfo {title} {On the {{Einstein Podolsky Rosen}} paradox},\ }\href {https://doi.org/10.1103/PhysicsPhysiqueFizika.1.195} {\bibfield  {journal} {\bibinfo  {journal} {Physics}\ }\textbf {\bibinfo {volume} {1}},\ \bibinfo {pages} {195} (\bibinfo {year} {1964})}\BibitemShut {NoStop}%
\bibitem [{\citenamefont {Bell}(1966)}]{bell1966problem}%
  \BibitemOpen
  \bibfield  {author} {\bibinfo {author} {\bibfnamefont {J.~S.}\ \bibnamefont {Bell}},\ }\bibfield  {title} {\bibinfo {title} {On the {{Problem}} of {{Hidden Variables}} in {{Quantum Mechanics}}},\ }\href {https://doi.org/10.1103/RevModPhys.38.447} {\bibfield  {journal} {\bibinfo  {journal} {Reviews of Modern Physics}\ }\textbf {\bibinfo {volume} {38}},\ \bibinfo {pages} {447} (\bibinfo {year} {1966})}\BibitemShut {NoStop}%
\bibitem [{\citenamefont {Ollitrault}\ \emph {et~al.}(2020)\citenamefont {Ollitrault}, \citenamefont {Kandala}, \citenamefont {Chen}, \citenamefont {Barkoutsos}, \citenamefont {Mezzacapo}, \citenamefont {Pistoia}, \citenamefont {Sheldon}, \citenamefont {Woerner}, \citenamefont {Gambetta},\ and\ \citenamefont {Tavernelli}}]{ollitrault2020quantum}%
  \BibitemOpen
  \bibfield  {author} {\bibinfo {author} {\bibfnamefont {P.~J.}\ \bibnamefont {Ollitrault}}, \bibinfo {author} {\bibfnamefont {A.}~\bibnamefont {Kandala}}, \bibinfo {author} {\bibfnamefont {C.-F.}\ \bibnamefont {Chen}}, \bibinfo {author} {\bibfnamefont {P.~K.}\ \bibnamefont {Barkoutsos}}, \bibinfo {author} {\bibfnamefont {A.}~\bibnamefont {Mezzacapo}}, \bibinfo {author} {\bibfnamefont {M.}~\bibnamefont {Pistoia}}, \bibinfo {author} {\bibfnamefont {S.}~\bibnamefont {Sheldon}}, \bibinfo {author} {\bibfnamefont {S.}~\bibnamefont {Woerner}}, \bibinfo {author} {\bibfnamefont {J.~M.}\ \bibnamefont {Gambetta}},\ and\ \bibinfo {author} {\bibfnamefont {I.}~\bibnamefont {Tavernelli}},\ }\bibfield  {title} {\bibinfo {title} {Quantum equation of motion for computing molecular excitation energies on a noisy quantum processor},\ }\href {https://doi.org/10.1103/PhysRevResearch.2.043140} {\bibfield  {journal} {\bibinfo  {journal} {Physical Review Research}\ }\textbf {\bibinfo {volume} {2}},\ \bibinfo {pages} {043140}
  (\bibinfo {year} {2020})}\BibitemShut {NoStop}%
\bibitem [{\citenamefont {Ralli}\ and\ \citenamefont {Weaving}(2022)}]{ralli2022symmer}%
  \BibitemOpen
  \bibfield  {author} {\bibinfo {author} {\bibfnamefont {A.}~\bibnamefont {Ralli}}\ and\ \bibinfo {author} {\bibfnamefont {T.}~\bibnamefont {Weaving}},\ }\href {https://github.com/UCL-CCS/symmer} {\bibinfo {title} {Symmer}} (\bibinfo {year} {2022})\BibitemShut {NoStop}%
\bibitem [{\citenamefont {Aleksandrowicz}\ \emph {et~al.}(2019)\citenamefont {Aleksandrowicz} \emph {et~al.}}]{gadi_aleksandrowicz_2019_2562111}%
  \BibitemOpen
  \bibfield  {author} {\bibinfo {author} {\bibfnamefont {G.}~\bibnamefont {Aleksandrowicz}} \emph {et~al.},\ }\href {https://doi.org/10.5281/zenodo.2562111} {\bibinfo {title} {Qiskit: {{An}} open-source framework for quantum computing}},\ \bibinfo {howpublished} {Zenodo} (\bibinfo {year} {2019})\BibitemShut {NoStop}%
\bibitem [{\citenamefont {McClean}\ \emph {et~al.}(2020)\citenamefont {McClean}, \citenamefont {Rubin}, \citenamefont {Sung}, \citenamefont {Kivlichan}, \citenamefont {{Bonet-Monroig}}, \citenamefont {Cao}, \citenamefont {Dai}, \citenamefont {Fried}, \citenamefont {Gidney}, \citenamefont {Gimby}, \citenamefont {Gokhale}, \citenamefont {H{\"a}ner}, \citenamefont {Hardikar}, \citenamefont {Havl{\'i}{\v c}ek}, \citenamefont {Higgott}, \citenamefont {Huang}, \citenamefont {Izaac}, \citenamefont {Jiang}, \citenamefont {Liu}, \citenamefont {McArdle}, \citenamefont {Neeley}, \citenamefont {O'Brien}, \citenamefont {O'Gorman}, \citenamefont {Ozfidan}, \citenamefont {Radin}, \citenamefont {Romero}, \citenamefont {Sawaya}, \citenamefont {Senjean}, \citenamefont {Setia}, \citenamefont {Sim}, \citenamefont {Steiger}, \citenamefont {Steudtner}, \citenamefont {Sun}, \citenamefont {Sun}, \citenamefont {Wang}, \citenamefont {Zhang},\ and\ \citenamefont {Babbush}}]{mcclean2020openfermion}%
  \BibitemOpen
  \bibfield  {author} {\bibinfo {author} {\bibfnamefont {J.~R.}\ \bibnamefont {McClean}}, \bibinfo {author} {\bibfnamefont {N.~C.}\ \bibnamefont {Rubin}}, \bibinfo {author} {\bibfnamefont {K.~J.}\ \bibnamefont {Sung}}, \bibinfo {author} {\bibfnamefont {I.~D.}\ \bibnamefont {Kivlichan}}, \bibinfo {author} {\bibfnamefont {X.}~\bibnamefont {{Bonet-Monroig}}}, \bibinfo {author} {\bibfnamefont {Y.}~\bibnamefont {Cao}}, \bibinfo {author} {\bibfnamefont {C.}~\bibnamefont {Dai}}, \bibinfo {author} {\bibfnamefont {E.~S.}\ \bibnamefont {Fried}}, \bibinfo {author} {\bibfnamefont {C.}~\bibnamefont {Gidney}}, \bibinfo {author} {\bibfnamefont {B.}~\bibnamefont {Gimby}}, \bibinfo {author} {\bibfnamefont {P.}~\bibnamefont {Gokhale}}, \bibinfo {author} {\bibfnamefont {T.}~\bibnamefont {H{\"a}ner}}, \bibinfo {author} {\bibfnamefont {T.}~\bibnamefont {Hardikar}}, \bibinfo {author} {\bibfnamefont {V.}~\bibnamefont {Havl{\'i}{\v c}ek}}, \bibinfo {author} {\bibfnamefont {O.}~\bibnamefont {Higgott}}, \bibinfo {author}
  {\bibfnamefont {C.}~\bibnamefont {Huang}}, \bibinfo {author} {\bibfnamefont {J.}~\bibnamefont {Izaac}}, \bibinfo {author} {\bibfnamefont {Z.}~\bibnamefont {Jiang}}, \bibinfo {author} {\bibfnamefont {X.}~\bibnamefont {Liu}}, \bibinfo {author} {\bibfnamefont {S.}~\bibnamefont {McArdle}}, \bibinfo {author} {\bibfnamefont {M.}~\bibnamefont {Neeley}}, \bibinfo {author} {\bibfnamefont {T.}~\bibnamefont {O'Brien}}, \bibinfo {author} {\bibfnamefont {B.}~\bibnamefont {O'Gorman}}, \bibinfo {author} {\bibfnamefont {I.}~\bibnamefont {Ozfidan}}, \bibinfo {author} {\bibfnamefont {M.~D.}\ \bibnamefont {Radin}}, \bibinfo {author} {\bibfnamefont {J.}~\bibnamefont {Romero}}, \bibinfo {author} {\bibfnamefont {N.~P.~D.}\ \bibnamefont {Sawaya}}, \bibinfo {author} {\bibfnamefont {B.}~\bibnamefont {Senjean}}, \bibinfo {author} {\bibfnamefont {K.}~\bibnamefont {Setia}}, \bibinfo {author} {\bibfnamefont {S.}~\bibnamefont {Sim}}, \bibinfo {author} {\bibfnamefont {D.~S.}\ \bibnamefont {Steiger}}, \bibinfo {author} {\bibfnamefont
  {M.}~\bibnamefont {Steudtner}}, \bibinfo {author} {\bibfnamefont {Q.}~\bibnamefont {Sun}}, \bibinfo {author} {\bibfnamefont {W.}~\bibnamefont {Sun}}, \bibinfo {author} {\bibfnamefont {D.}~\bibnamefont {Wang}}, \bibinfo {author} {\bibfnamefont {F.}~\bibnamefont {Zhang}},\ and\ \bibinfo {author} {\bibfnamefont {R.}~\bibnamefont {Babbush}},\ }\bibfield  {title} {\bibinfo {title} {{{OpenFermion}}: The electronic structure package for quantum computers},\ }\href {https://doi.org/10.1088/2058-9565/ab8ebc} {\bibfield  {journal} {\bibinfo  {journal} {Quantum Science and Technology}\ }\textbf {\bibinfo {volume} {5}},\ \bibinfo {pages} {034014} (\bibinfo {year} {2020})}\BibitemShut {NoStop}%
\bibitem [{\citenamefont {Anand}\ \emph {et~al.}(2022)\citenamefont {Anand}, \citenamefont {Schleich}, \citenamefont {{Alperin-Lea}}, \citenamefont {Jensen}, \citenamefont {Sim}, \citenamefont {{D{\'i}az-Tinoco}}, \citenamefont {Kottmann}, \citenamefont {Degroote}, \citenamefont {Izmaylov},\ and\ \citenamefont {{Aspuru-Guzik}}}]{anand2022quantum}%
  \BibitemOpen
  \bibfield  {author} {\bibinfo {author} {\bibfnamefont {A.}~\bibnamefont {Anand}}, \bibinfo {author} {\bibfnamefont {P.}~\bibnamefont {Schleich}}, \bibinfo {author} {\bibfnamefont {S.}~\bibnamefont {{Alperin-Lea}}}, \bibinfo {author} {\bibfnamefont {P.~W.~K.}\ \bibnamefont {Jensen}}, \bibinfo {author} {\bibfnamefont {S.}~\bibnamefont {Sim}}, \bibinfo {author} {\bibfnamefont {M.}~\bibnamefont {{D{\'i}az-Tinoco}}}, \bibinfo {author} {\bibfnamefont {J.~S.}\ \bibnamefont {Kottmann}}, \bibinfo {author} {\bibfnamefont {M.}~\bibnamefont {Degroote}}, \bibinfo {author} {\bibfnamefont {A.~F.}\ \bibnamefont {Izmaylov}},\ and\ \bibinfo {author} {\bibfnamefont {A.}~\bibnamefont {{Aspuru-Guzik}}},\ }\bibfield  {title} {\bibinfo {title} {A quantum computing view on unitary coupled cluster theory},\ }\href {https://doi.org/10.1039/D1CS00932J} {\bibfield  {journal} {\bibinfo  {journal} {Chemical Society Reviews}\ }\textbf {\bibinfo {volume} {51}},\ \bibinfo {pages} {1659} (\bibinfo {year} {2022})}\BibitemShut {NoStop}%
\bibitem [{\citenamefont {{Greene-Diniz}}\ and\ \citenamefont {Ramo}(2020)}]{greene-diniz2020generalized}%
  \BibitemOpen
  \bibfield  {author} {\bibinfo {author} {\bibfnamefont {G.}~\bibnamefont {{Greene-Diniz}}}\ and\ \bibinfo {author} {\bibfnamefont {D.~M.}\ \bibnamefont {Ramo}},\ }\bibfield  {title} {\bibinfo {title} {Generalized unitary coupled cluster excitations for multireference molecular states optimized by the variational quantum eigensolver},\ }\bibfield  {journal} {\bibinfo  {journal} {International Journal of Quantum Chemistry}\ }\href {https://doi.org/10.1002/qua.26352} {10.1002/qua.26352} (\bibinfo {year} {2020})\BibitemShut {NoStop}%
\bibitem [{\citenamefont {Qianjun}(2025)}]{yao2025csvqd}%
  \BibitemOpen
  \bibfield  {author} {\bibinfo {author} {\bibfnamefont {Y.}~\bibnamefont {Qianjun}},\ }\href {https://github.com/kkyoyoo/CS-VQD-with-Symm-Opt} {\bibinfo {title} {Cs-vqd with symmetry optimize}} (\bibinfo {year} {2025})\BibitemShut {NoStop}%
\bibitem [{\citenamefont {Schultz}\ \emph {et~al.}(2004)\citenamefont {Schultz}, \citenamefont {Samoylova}, \citenamefont {Radloff}, \citenamefont {Hertel}, \citenamefont {Sobolewski},\ and\ \citenamefont {Domcke}}]{schultz2004efficient}%
  \BibitemOpen
  \bibfield  {author} {\bibinfo {author} {\bibfnamefont {T.}~\bibnamefont {Schultz}}, \bibinfo {author} {\bibfnamefont {E.}~\bibnamefont {Samoylova}}, \bibinfo {author} {\bibfnamefont {W.}~\bibnamefont {Radloff}}, \bibinfo {author} {\bibfnamefont {I.~V.}\ \bibnamefont {Hertel}}, \bibinfo {author} {\bibfnamefont {A.~L.}\ \bibnamefont {Sobolewski}},\ and\ \bibinfo {author} {\bibfnamefont {W.}~\bibnamefont {Domcke}},\ }\bibfield  {title} {\bibinfo {title} {Efficient {{Deactivation}} of a {{Model Base Pair}} via {{Excited-State Hydrogen Transfer}}},\ }\href {https://doi.org/10.1126/science.1104038} {\bibfield  {journal} {\bibinfo  {journal} {Science}\ }\textbf {\bibinfo {volume} {306}},\ \bibinfo {pages} {1765} (\bibinfo {year} {2004})}\BibitemShut {NoStop}%
\bibitem [{\citenamefont {Leone}\ \emph {et~al.}(2024)\citenamefont {Leone}, \citenamefont {Oliviero}, \citenamefont {Cincio},\ and\ \citenamefont {Cerezo}}]{leone2024practical}%
  \BibitemOpen
  \bibfield  {author} {\bibinfo {author} {\bibfnamefont {L.}~\bibnamefont {Leone}}, \bibinfo {author} {\bibfnamefont {S.~F.~E.}\ \bibnamefont {Oliviero}}, \bibinfo {author} {\bibfnamefont {L.}~\bibnamefont {Cincio}},\ and\ \bibinfo {author} {\bibfnamefont {M.}~\bibnamefont {Cerezo}},\ }\bibfield  {title} {\bibinfo {title} {On the practical usefulness of the {{Hardware Efficient Ansatz}}},\ }\href {https://doi.org/10.22331/q-2024-07-03-1395} {\bibfield  {journal} {\bibinfo  {journal} {Quantum}\ }\textbf {\bibinfo {volume} {8}},\ \bibinfo {pages} {1395} (\bibinfo {year} {2024})}\BibitemShut {NoStop}%
\bibitem [{\citenamefont {McClean}\ \emph {et~al.}(2018)\citenamefont {McClean}, \citenamefont {Boixo}, \citenamefont {Smelyanskiy}, \citenamefont {Babbush},\ and\ \citenamefont {Neven}}]{mcclean2018barren}%
  \BibitemOpen
  \bibfield  {author} {\bibinfo {author} {\bibfnamefont {J.~R.}\ \bibnamefont {McClean}}, \bibinfo {author} {\bibfnamefont {S.}~\bibnamefont {Boixo}}, \bibinfo {author} {\bibfnamefont {V.~N.}\ \bibnamefont {Smelyanskiy}}, \bibinfo {author} {\bibfnamefont {R.}~\bibnamefont {Babbush}},\ and\ \bibinfo {author} {\bibfnamefont {H.}~\bibnamefont {Neven}},\ }\bibfield  {title} {\bibinfo {title} {Barren plateaus in quantum neural network training landscapes},\ }\href {https://doi.org/10.1038/s41467-018-07090-4} {\bibfield  {journal} {\bibinfo  {journal} {Nature Communications}\ }\textbf {\bibinfo {volume} {9}},\ \bibinfo {pages} {4812} (\bibinfo {year} {2018})}\BibitemShut {NoStop}%
\bibitem [{\citenamefont {Mao}\ \emph {et~al.}(2024)\citenamefont {Mao}, \citenamefont {Tian},\ and\ \citenamefont {Sun}}]{mao2024determining}%
  \BibitemOpen
  \bibfield  {author} {\bibinfo {author} {\bibfnamefont {R.}~\bibnamefont {Mao}}, \bibinfo {author} {\bibfnamefont {G.}~\bibnamefont {Tian}},\ and\ \bibinfo {author} {\bibfnamefont {X.}~\bibnamefont {Sun}},\ }\bibfield  {title} {\bibinfo {title} {Towards determining the presence of barren plateaus in some chemically inspired variational quantum algorithms},\ }\href {https://doi.org/10.1038/s42005-024-01798-0} {\bibfield  {journal} {\bibinfo  {journal} {Communications Physics}\ }\textbf {\bibinfo {volume} {7}},\ \bibinfo {pages} {342} (\bibinfo {year} {2024})}\BibitemShut {NoStop}%
\bibitem [{\citenamefont {Lyu}\ \emph {et~al.}(2023)\citenamefont {Lyu}, \citenamefont {Xu}, \citenamefont {Yung},\ and\ \citenamefont {Bayat}}]{lyu2023symmetry}%
  \BibitemOpen
  \bibfield  {author} {\bibinfo {author} {\bibfnamefont {C.}~\bibnamefont {Lyu}}, \bibinfo {author} {\bibfnamefont {X.}~\bibnamefont {Xu}}, \bibinfo {author} {\bibfnamefont {M.-H.}\ \bibnamefont {Yung}},\ and\ \bibinfo {author} {\bibfnamefont {A.}~\bibnamefont {Bayat}},\ }\bibfield  {title} {\bibinfo {title} {Symmetry enhanced variational quantum spin eigensolver},\ }\href {https://doi.org/10.22331/q-2023-01-19-899} {\bibfield  {journal} {\bibinfo  {journal} {Quantum}\ }\textbf {\bibinfo {volume} {7}},\ \bibinfo {pages} {899} (\bibinfo {year} {2023})}\BibitemShut {NoStop}%
\bibitem [{\citenamefont {Vatan}\ and\ \citenamefont {Williams}(2004)}]{vatan2004optimal}%
  \BibitemOpen
  \bibfield  {author} {\bibinfo {author} {\bibfnamefont {F.}~\bibnamefont {Vatan}}\ and\ \bibinfo {author} {\bibfnamefont {C.}~\bibnamefont {Williams}},\ }\bibfield  {title} {\bibinfo {title} {Optimal quantum circuits for general two-qubit gates},\ }\href {https://doi.org/10.1103/PhysRevA.69.032315} {\bibfield  {journal} {\bibinfo  {journal} {Physical Review A}\ }\textbf {\bibinfo {volume} {69}},\ \bibinfo {pages} {032315} (\bibinfo {year} {2004})}\BibitemShut {NoStop}%
\end{thebibliography}%

\end{document}